\documentstyle[sprocl,graphicx]{article}

\def\be{\begin{equation}}
\def\ee{\end{equation}}
\def\bea{\begin{eqnarray}}
\def\eea{\end{eqnarray}}

\newcommand{\cent}[1] {\begin{center}#1\end{center}}
\newcommand{\doublint}{\int\rule{-3.5mm}{0mm}\int} 

\newcommand{\vecbm}[1]{\mbox{$\boldmath#1$}}

\newcommand{\lra} {$\leftrightarrow$} 
\newcommand{\vecb}[1]{\mbox{\bf#1}}
\begin{document}

\title{What can nuclear collisions teach us about the boiling of
  water\\or the formation of multi-star systems ?\footnote{Invited
    talk Bologna 2000 - Structure of the Nucleus at the Dawn of the
    Century, Bologna, Italia May 29 - June 3, 2000}} \author{D.H.E.
    Gross}

\address{Hahn-Meitner-Institut
  Berlin, Bereich Theoretische Physik,Glienickerstr.100\\ 14109
  Berlin, Germany and Freie Universit{\"a}t Berlin, Fachbereich
  Physik}


\maketitle\abstracts{Phase transitions in nuclei, small atomic
  clusters and self-gravitating systems demand the extension of
  thermo-statistics to ``Small'' systems. The main obstacle is the
  thermodynamic limit. It is shown how the original definition of the
  entropy by Boltzmann as the volume of the energy-manifold of the
  N-body phase space allows a {\em geometrical} definition of the
  entropy as function of the conserved quantities.  Without invoking
  the thermodynamic limit the whole ``zoo'' of phase transitions and
  critical points/lines can be unambiguously defined.  The relation to
  the Yang--Lee singularities of the grand-canonical partition sum is
  pointed out. It is shown that just phase transitions in
  non-extensive systems give the complete set of characteristic
  parameters of the transition {\em including the surface tension.}
  Nuclear heavy-ion collisions are an experimental playground to
  explore this extension of thermo-statistics}
\noindent Thermodynamicists will certainly answer our question by:
``Nothing'' and yet:
\section{There is a lot to add to classical equilibrium
  statistics from our experience with ``Small'' systems:} Following
Lieb \cite{lieb97} extensivity \footnote{Dividing extensive systems
  into larger pieces, the total energy and entropy are equal to the
  sum of those of the pieces.} and the existence of the thermodynamic
limit $N\to\infty|_{N/V= const}$ are essential conditions for
conventional (canonical) thermodynamics to apply.  Certainly, this
implies also the homogeneity of the system.  Phase transitions are
somehow foreign to this: The essence of first order transitions is
that the systems become inhomogeneous and split into different phases
separated by interfaces. In general are phase transitions consequently
represented by singularities in the grand-canonical partition sum
(Yang--Lee singularities). In the following we show that the
micro-canonical ensemble gives much more detailed insight.

There is a whole group of physical many-body systems called ``Small''
in the following which cannot be addressed by conventional
thermo-statistics:
\begin{itemize}
\item nuclei,
\item atomic cluster
\item polymers
\item soft matter (biological) systems
\item astrophysical systems
\item first order transitions are distinguished from continuous
  transitions by the appearance of phase-separations and interfaces
  with surface tension. If the range of the force or the thickness of
  the surface layers is such that the number of surface particles is
  not negligible compared to the total number of particles, these
  systems are non-extensive.
\end{itemize}

It is common to all these examples, that the systems are
non-extensive. For such systems the thermodynamic limit does not exist
or makes no sense. Either the range of the forces (Coulomb,
gravitation) are of the order of the linear dimensions of these
systems, and/or they are strongly inhomogeneous e.g. at
phase-separation. Before inventing any new type of non-extensive
thermodynamics or entropy like the one introduced by Tsallis
\cite{tsallis88} we should realize that Boltzmann's definition:\\
\cent{\fbox{\fbox{\vecbm{S=k*lnW}}}} with
\begin{eqnarray}
W(E,N,V)&=&\epsilon_0 tr\delta(E-H_N)\\
tr\delta(E-H_N)&=&\int{\frac{d^{3N}p\;d^{3N}q}{N!(2\pi\hbar)^{3N}}
\delta(E-H_N)}.\nonumber
\end{eqnarray} ($\epsilon_0$ a suitable small
energy constant) does not invoke the thermodynamic limit, nor
additivity, nor extensivity, nor concavity of the entropy $S(E,N)$
(downwards bending).  This was largely forgotten since hundred years.
We have to go back to pre Gibbsian times. It is a purely geometrical
definition of the entropy and applies as well to ``Small'' systems.
Moreover, the entropy $S(E,N)$ as defined above {\em is everywhere
  single-valued and multiple differentiable.}  There are no
singularities in it.  This is the most simple access to equilibrium
statistics. We will explore its consequences in this contribution.
Moreover, we will see that this way we get {\em simultaneously the
  complete} information about the three crucial parameters
characterizing a phase transition of first order: transition
temperature $T_{tr}$, latent heat per atom $q_{lat}$ and surface
tension $\sigma_{surf}$. Boltzmann's famous epitaph above contains
everything what can be said about equilibrium thermodynamics in its
most condensed form. $W$ is the volume of the sub-manifold at sharp
energy in the $6N$-dim. phase space.
\section{Relation of the topology of $S(E,N)$ to the Yang-Lee 
  singularities} In conventional thermo-statistics phase transitions
are indicated by singularities of the grand-canonical partition
function $Z(T,\mu,V)$, $V$ is the volume. See more details in
\cite{gross174,gross173,gross175,gross176}
\begin{eqnarray}
 Z(T,\mu,V)&=&\doublint_0^{\infty}{\frac{dE}{\epsilon_0}\;dN\;e^{-[E-\mu
N-TS(E)]/T}}\nonumber\\
&&=\frac{V^2}{\epsilon_0}\doublint_0^{\infty}{de\;dn\;e^{-V[e-\mu
n-T{\mbox{\boldmath$\scriptstyle s(e,n)$}}]/T}}\label{grandsum}.\\
&\approx&\hspace{2cm}e^{\mbox{ const.+lin.+quadr.}}\nonumber
\end{eqnarray}
in the thermodynamic limit $V\to\infty|_{N/V= const}$.

The double Laplace integral (\ref{grandsum}) can be evaluated
asymptotically for large $V$ by expanding the exponent as indicated in
the last line to second order in $\Delta e,\Delta n$ around the
``stationary point'' $e_s,n_s$ where the linear term vanish:
\begin{eqnarray}
\frac{1}{T}&=&\left.\frac{\partial S}{\partial E}\right|_s\nonumber\\
\frac{\mu}{T}&=&-\left.\frac{\partial S}{\partial N}\right|_s\nonumber\\
\frac{P}{T}&=&\left.\frac{\partial S}{\partial V}\right|_s
\end{eqnarray}
the only term remaining to be integrated is the quadratic one. {\em If
the two eigen-curvatures $\lambda_1<0$, $\lambda_2<0$} this is then a
Gaussian integral and yields:
\begin{eqnarray}
Z(T,\mu,V)&=&\frac{V^2}{\epsilon_0}e^{-V[e_s-\mu 
n_s-T{\mbox{\boldmath$\scriptstyle s(e_s,n_s)$}}]/T}}
\doublint_0^{\infty}{dv_1\;dv_2\;
e^{V[\lambda_1 v_1^2+\lambda_2 v_2^2]/2}\\
Z(T,\mu,V)&=&e^{-F(T,\mu,V)}\\
\frac{F(T,\mu,V)}{V}&\to& e_s-\mu n_s-Ts_s
+\frac{T\ln{(\sqrt{\det(e_s,n_s)})}}{V}+o(\frac{\ln{V}}{V})\label{asympt}.
\end{eqnarray}
Here $\det(e_s,n_s)$ is the determinant of the {\em curvatures} of $s(e,n)$.
\begin{equation}
\det(e,n)= \left\|\begin{array}{cc}
\frac{\partial^2 s}{\partial e^2}& \frac{\partial^2 s}{\partial n\partial e}\\
\frac{\partial^2 s}{\partial e\partial n}& \frac{\partial^2 s}{\partial n^2}
\end{array}\right\|
= \left\|\begin{array}{cc}
s_{ee}&s_{en}\\
s_{ne}&s_{nn}
\end{array}\right\|=\lambda_1\lambda_2,\hspace{1cm}\lambda_1\ge\lambda_2
 \label{curvdet}
\end{equation}
In the cases studied here $\lambda_2<0$ but $\lambda_1$ can be positive
or negative. If $\det(e_s,n_s)$ is positive ($\lambda_1<0$) the last
two terms in eq.(\ref{asympt}) go to $0$, and we obtain the familiar
result $f(T,\mu,V)=e_s-\mu n_s-Ts_s$. I.e.  the {\em curvature
  $\lambda_1$ of the entropy surface $s(e,n,V)$ decides whether the
  grand-canonical ensemble agrees with the fundamental micro ensemble
  in the thermodynamic limit.} If this is the case, $Z(T,\mu,V)$ is
analytical and due to Yang and Lee we have a single, stable phase. Or
otherwise, {\em the Yang-Lee singularities reflect anomalous
  points/regions of $\lambda_1\ge 0$} ($\det(e,n)\le 0$). This is
crucial.  As $\det(e_s,n_s)$ can be studied for finite or even small
systems as well, this is the only proper extension of phase
transitions to ``Small'' systems.
\section{The physical origin of the wrong (positive) curvature
$\lambda_1$ of $s(e_s,n_s)$} Before we proceed with the general
micro-canonical classification of the various types of phase
transitions we will now discuss the physical origin of convex
(upwards bending) intruders in the entropy surface.

As nuclear matter is not accessible to us we should investigate atomic
clusters and compare their thermodynamic behavior with that of the 
known bulk. 

\begin{figure}
\cent{\includegraphics*[bb = 99 57 400 286, angle=-0, width=6cm,  
clip=true]{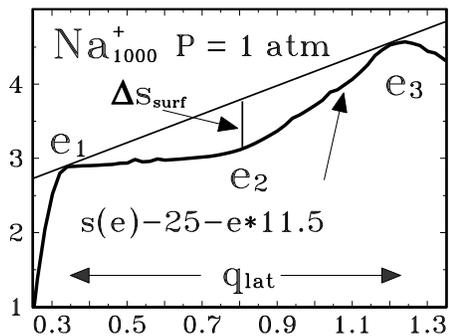}}
\caption{MMMC simulation of the entropy $s(e)$ ($e$ in eV per atom)
of a system of $1000$ sodium atoms with realistic interaction at an
external pressure of 1 atm.  At the energy per atom $e_1$ the system
is in the pure liquid phase and at $e_3$ in the pure gas phase, of
course with fluctuations. The latent heat per atom is
$q_{lat}=e_3-e_1$.  Attention: the curve $s(e)$ is artifically
sheared by subtracting a linear function $25+e*11.5$ in order to make
the convex intruder visible.  {\em $s(e)$ is always a steeply
monotonic rising function.} We clearly see the global concave
(downwards bending) nature of $s(e)$ and its convex intruder. Its depth
is the entropy loss due to the additional correlations by the
interfaces.  From this one can calculate the surface tension
$\sigma_{surf}/T_{tr}=\Delta s_{surf}*N/N_{surf}$. The double tangent
is the concave hull of $s(e)$. Its derivative gives the Maxwell line
in the caloric curve $T(e)$ at $T_{tr}$}\label{naprl0}
\end{figure}

In the figure (\ref{naprl0}) we compare the ``liquid--gas'' phase
transition in sodium clusters of a few hundred atoms with that of the
bulk at 1 atm..

In the table (\ref{tab}) we show in comparison with the known bulk
values the four important parameters, transition temperature $T_{tr}$,
latent heat per atom $q_{lat}$, the entropy gain for the evaporation
of one atom $s_{boil}$ as proposed by the empirical Trouton's rule
($\sim 10$) and $\Delta s_{surf}$, the surface entropy per atom as
defined above.  $N^{2/3}_{eff}$ is the average number of surface atoms
of the clusters. $\sigma/T_{tr}$ is the surface tension over the
transition temperature.

\begin{table}\cent{
\begin{tabular} {|c|c|c|c|c|c|} \hline 
&$N_0$&$200$&$1000$&$3000$&\vecb{bulk}\\ 
\hline 
\hline  
&$T_{tr} \;[K]$&$940$&$990$&$1095$&\vecb{1156}\\ \cline{2-6} 
&$q_{lat} \;[eV]$&$0.82$&$0.91$&$0.94$&\vecb{0.923}\\ \cline{2-6} 
{\bf Na}&$s_{boil}$&$10.1$&$10.7$&$9.9$&\vecb{9.267}\\ \cline{2-6} 
&$\Delta s_{surf}$&$0.55$&$0.56$&$0.45$&\\ \cline{2-6} 
&$N_{eff}^{2/3}$&$39.94$&$98.53$&$186.6$&$\vecbm{\infty}$\\
\cline{2-6} 
&$\sigma/T_{tr}$&$2.75$&$5.68$&$7.07$&\vecb{7.41}\\ 
\hline
\end{tabular}\caption{Parameters of the liquid--gas transition
of small sodium clusters (MMMC-calculation) in comparison with 
the bulk.}\label{tab}}
\end{table}
\section{Phase transitions in the micro-ensemble:
  The topology of the determinant of curvatures of
  {\boldmath$s(e,n)$:}} Now we can give a systematic and generic
classification of phase transitions which applies also to ``Small''
systems and their relation to the Yang-Lee singularities:
\begin{itemize}
\item \begin{minipage}[t]{5.5cm} A {\bf single stable} phase by
    $d(e,n)>0$ ($\lambda_1<0$). Here $s(e,n)$ is concave (downwards
    bended) in both directions. Then there is a one to one mapping of
    the grand-canonical \lra the micro-ensemble. In the two examples
    on the right the order parameter, the direction $\vecb{v}_1$ of
    the eigenvector of largest curvature $\lambda_1$ was simply 
    assumed to be the energy.
\end{minipage}~\begin{minipage}[t]{6cm} \vspace{-0.5cm}
\includegraphics*[bb = 117 13 490 645, angle=-90, width=5.5cm,
clip=true]{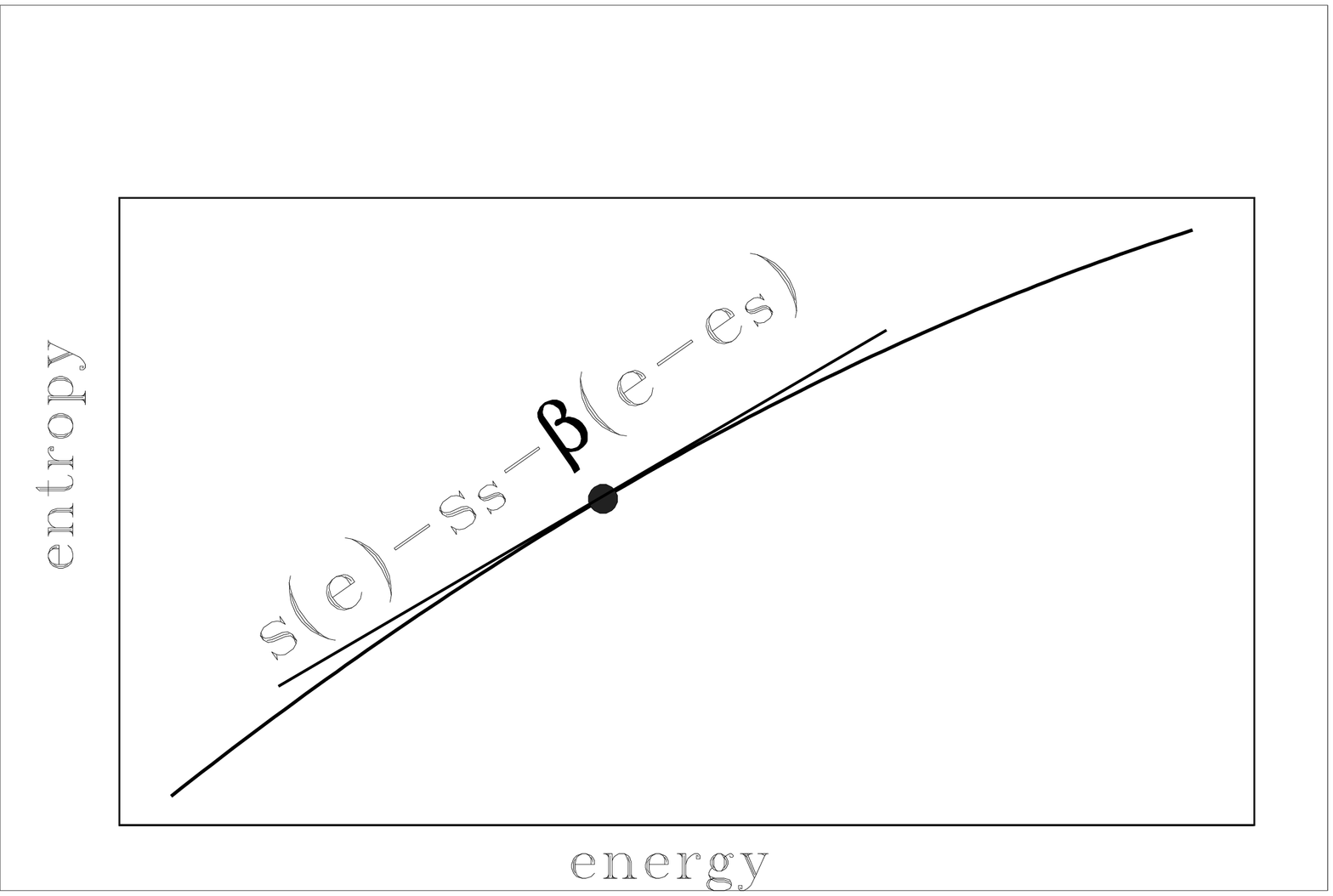}\end{minipage}
\item \begin{minipage}[t]{5.5cm} A {\bf transition of first order} with
    phase separation and surface tension is indicated by $d(e,n)<0$
    ($\lambda_1>0$). $s(e,n)$ has a convex intruder (upwards bended)
    in the direction $\vecb{v}_1$ of the largest curvature.
    The whole convex area of \{e,n\} is
    mapped into a single point in the canonical
    ensemble.\label{convex}.  I.e. if the curvature of $S(E,N)$ is
    $\lambda_1\ge 0$ {\bf both ensembles are not equivalent.}
\end{minipage}~\begin{minipage}[t]{6cm} \includegraphics*[bb = 130 8 490 640, angle=-90, width=5.5cm, clip=true]{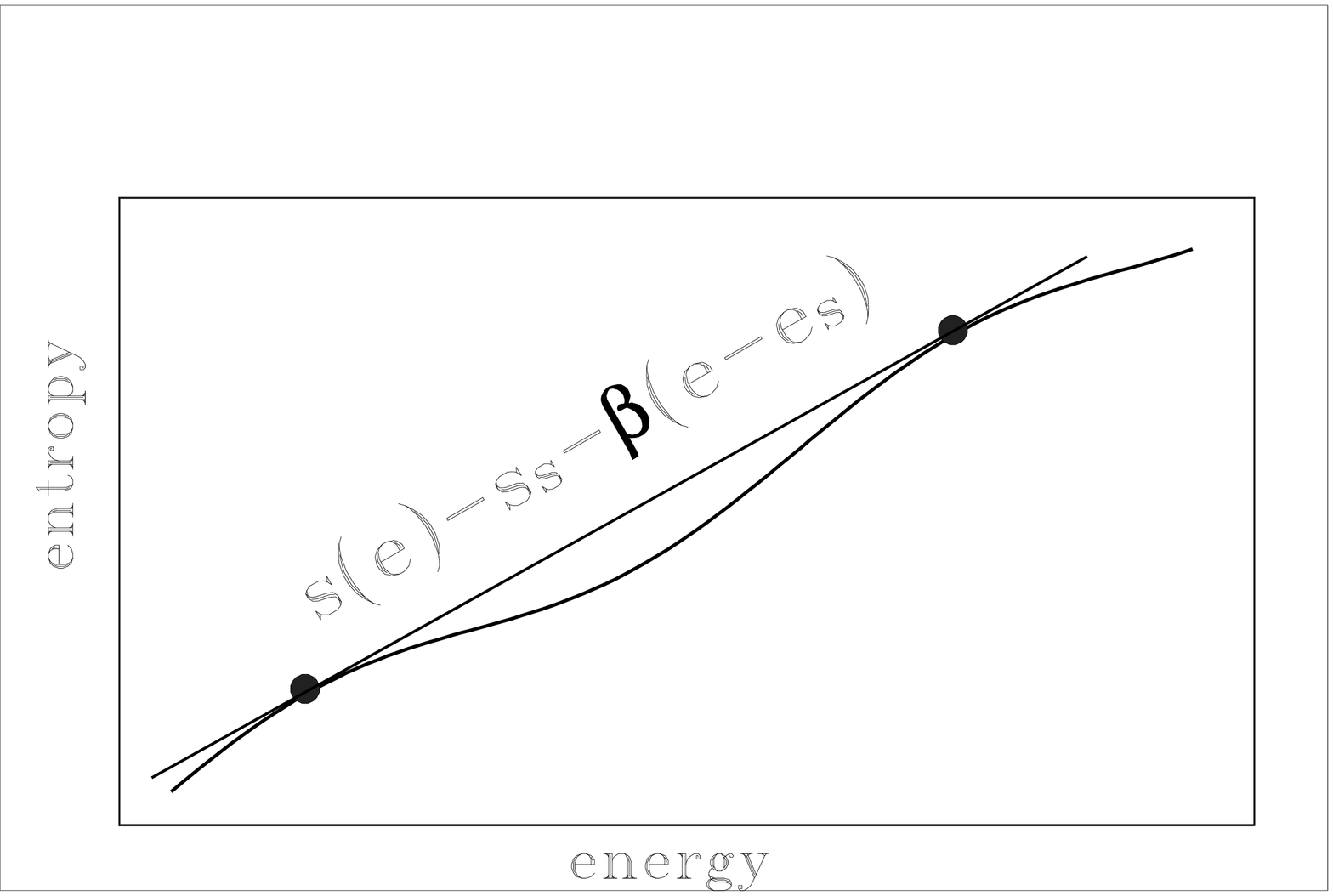} \end{minipage}
\item A {\bf continuous (``second order'')} transition with vanishing
  surface tension, where two neighboring phases become
  indistinguishable. This is at points where the two stationary points
  move into one another. I.e. where $d(e,n)=0$ and
  $\vecb{v}_{\lambda=0}\cdot\vecbm{\nabla}d=0$.  These are the {\em
    catastrophes} of the Laplace transform $E\to T$. Here
  $\vecb{v}_{\lambda=0}$ is the eigenvector of $d$ belonging to the
  largest curvature eigenvalue $\lambda=0$ ($\to$ order parameter).
\item Finally, a {\bf multi-critical point} where more than two phases
  become indistinguishable is at the branching of several lines in the
  \{$e,n$\}-phase-diagram with $d=0$, {\boldmath$\vecbm{\nabla}
    d=0$}.
\end{itemize}
  An example showing all these possible types of phase transitions in
  a small system is discussed in \cite{gross174,gross173,gross175}
\section{On the statistical formation of multi-star systems under rotation}
Having discussed the micro-canonical description of phase transitions
in ``Small'' systems we sketch here the relevance of our new
formulation of thermo-statistics for astrophysical systems. A finite
self-gravitating system is controlled by its accessible phase space
and its topology:
\begin{itemize}
\item Due to the long-range gravity the system is non-extensive.
\item Because there is no heat-, no angular momentum bath a microcanonical
treatment under the observation of the conservation laws is neccessary.
\item Attractive gravity leads to a collaps (phase transition of first
order).
\item Under rotation the collapsed system may multi-fragment and
  multi-star systems will be formed.
\item Also the fragmentation of Shoemaker-Levy 9 may be viewed as
  statistical multifragmentation under the linear stress of Jupiter's
  gravitation field.
\end{itemize}
With the analogy to the long-range Coulomb force one may
study many aspects by the study of collision of heavy nuclei. E.g.:
\begin{itemize}
\item At higher rotation the system prefers larger moment of inertia.
  This leads to more symmetric heaviest fragments.  For the
  astro-problem higher angular momentum leads to double or multiple
  stars instead of a mono star.
\item For heavy-ion collisions, rotation leads to higher radial energy
  of fragments, larger variance in the statistical fragmentation which
  must be distinguished from from radial flow !
\end{itemize} 
\section{Conclusion}
Instead of using the boiling of water we used the boiling of sodium to
demonstrate the physics of first order phase transitions and
especially the surface tension. ``Small'' systems show clearly how to
determine {\em all} important characteristica of first order
transitions: transition temperature, latent heat, {\em and surface
  tension}. This is possible just {\em because these systems are not
extensive and their entropy $s(e,n)$ has convex intruders from which
the surface entropy can be determined.} In the thermodynamic limit
this fact becomes more obscured than illuminated.

We also mentioned collapsing of self-gravitating and rotating hydrogen
clouds towards multi-star systems. As far as the structure of the
phase space is concerned there is a striking analogy to the phase
space of fragmenting hot nuclei under rotation. The experimental study
of the latter is also interesting in view of this analogy. One of the
main differences is of course that the latter are experimentally
accessible. Another important question is whether there is a kind of
statistical equilibrium established in the astrophysical systems and
if what the equilibration mechanism might be. Here our learning
process has just started.


\end{document}